\documentclass{article}

\usepackage{arxiv}

\usepackage[T1]{fontenc}    
\usepackage[utf8]{inputenc} 
\usepackage{lmodern}
\usepackage{hyperref}       
\usepackage{url}            
\usepackage{booktabs}       
\usepackage{amsfonts}       
\usepackage{nicefrac}       
\usepackage{microtype}      
\usepackage{lipsum}
\usepackage{graphicx}
\usepackage{amsmath}

\title{Individual Claims Forecasting with Bayesian Mixture Density Networks}

\author{
  Kevin Kuo \\
  Kasa AI\\
  \texttt{kevin@kasa.ai} \\
}

\begin{document}
\maketitle

\begin{abstract}
We introduce an individual claims forecasting framework utilizing Bayesian 
mixture density networks that can be used for claims analytics tasks such as
case reserving and triaging. The proposed approach enables incorporating claims
information from both structured and unstructured data sources, producing 
multi-period cash flow forecasts, and generating different scenarios of future 
payment patterns. We implement and evaluate the modeling framework using 
publicly available data.
\end{abstract}


\section{Introduction}

Individual claims reserving has garnered increasing interest in recent years.
While the main benefit cited for performing reserving at the claim level over 
aggregate loss triangle approaches, such as chain ladder and 
Bornhuetter-Ferguson, is potential improvement in predictive accuracy, 
especially in environments with changing portfolio mix 
\cite{boumezouedIndividualClaims2017}, there are additional practical advantages
to forecasting individual claim behavior. These include being able to obtain
updated views of portfolio risk as claims are reported and optimize adjuster 
resource allocation based on severity predictions. Although the benefits of 
individual claims modeling are promising, it has not yet achieved widespread 
adoption in practice. One contributing reason for the lack of adoption, we 
hypothesize, is the absence of a modeling framework with features important to
practitioners. We suggest that an effective loss reserves modeling framework 
should be able to:

\begin{itemize}
	\item Incorporate arbitrary claims information as predictors,
	\item Produce multi-period forecasts that are sufficiently stable over time, 
	and
	\item Sample different realizations of future payment patterns that encompass 
	both process uncertainty and model risk.
\end{itemize}
Companies are capturing increasingly diverse data, such as unstructured text 
from claims adjusters' notes and photographs of damages, that can potentially 
be predictive. Timing of cash flows and being able to sample from different 
future states have both business decision making and regulatory applications. A 
desirable characteristic of the forecasts is that they are stable over time, as 
management is averse to volatility in loss reserves figures from one accounting 
period to the next. A subjective criterion not listed above, which may affect
adoption of an approach, is that the models should be implementable without the
need for extensive bespoke feature engineering or specification of complex 
assumptions for underlying stochastic processes.

To the best of our knowledge, no existing loss reserving framework implements 
all of the above features. In this paper, we propose an extensible individual 
claims forecasting framework towards satisfying many of these criteria, utilizing
ideas from Bayesian neural networks (BNN) \cite{nealBayesianLearning2012} and 
mixture density networks (MDN) \cite{bishopMixtureDensity1994}. While we discuss
these concepts in detail later in the paper, at a high level,

\begin{itemize}
  \item BNNs are non-linear supervised learning models that capture complex 
  interactions among inputs, with prior distributions on model parameters; and
  \item MDNs are mixture models for conditional densities, where the mixture
  model parameters are the outputs of the neural network.
\end{itemize}
Concretely, our contributions are:

\begin{itemize}
	\item Development of an individual claims forecasting framework based on 
	Bayesian Mixture Density Networks (BMDN).
	\item Implementation of the proposed framework using publicly available 
	claims-level data, which provides a baseline for future work to compare 
	against.
\end{itemize}

\section{Related Work}

Claims-level reserving is a fast-moving research area, and 
\cite{boumezouedIndividualClaims2017} and \cite{taylorLossReserving2019} provide
recent surveys. Many of the current works in the area utilize machine learning 
(ML) techniques. Wüthrich \cite{wuthrichMachineLearning2018} introduces using machine 
learning algorithms to incorporate diverse claims characteristics inputs. It 
demonstrates a simple model where regression trees are used to predict the
number of payments, and suggests extensions such as compound modeling to predict
severity and bootstrap simulation to obtain prediction uncertainty. More 
recently,  \cite{duvalIndividualLoss2019, lopezTreeBasedAlgorithm2019, 
baudryMachineLearning} also utilize tree-based techniques for individual claims
reserving. Another direction of research for the individual claims reserving 
problem are the generative approaches of, for example, 
\cite{antonioMicrolevelStochastic2014, pigeonIndividualLoss2013, 
pigeonIndividualLoss2014}. In this latter category of methodologies, a set of
distributional assumptions are posited for the different drivers of claims, 
such as time to the next payment and its amount. These distributions are fit to
the data, then, with the obtained parameters, the modeler is able to perform 
simulations of future development paths by sampling from the fitted 
distributions. While this approach provides a natural way to obtain samples of
future cash flow paths, the distributional assumptions may be too rigid in some 
cases. It is also difficult to incorporate individual claim characteristics; to 
differentiate among different characteristics, one would have to segment the 
claims and fit separate models to each group.

In formulating our framework, we also draw inspiration from machine learning 
approaches to aggregate triangle data, including 
\cite{gabrielliNeuralNetwork2019a, gabrielliNeuralNetwork2019}, which embed a
classical parametric loss reserving models into neural networks, and the 
DeepTriangle \cite{kuo2019deeptriangle} framework, whose neural network 
architecture we adapt for individual claims data.

\section{Preliminaries}

We begin with a description of the loss reserving problem then briefly discuss 
BNN and MDN, two ideas that we incorporate into our claims forecasting neural 
network. As neural networks have been utilized and discussed extensively in 
recent loss reserving literature \cite{gabrielliNeuralNetwork2019a, 
gabrielliNeuralNetwork2019, kuo2019deeptriangle, wuthrichNeuralNetworks2018,
gabrielliIndividualClaims2018}, we defer discussion of neural network 
fundamentals to those works and the standard reference
\cite{goodfellowDeepLearning2016}. To aid the practicing actuary in consuming 
this paper, we expand on certain concepts when we introduce the proposed neural 
network architecture in Section \ref{sec:claims-forecasting-model}.

We remark that in sections \ref{section:bnn} and \ref{section:mdn}, we discuss 
Bayesian inference and mixture density networks, respectively, in general, and 
provide details on our specific choices of distributions later on in the paper.

\subsection{The Loss Reserving Problem}

\begin{figure}
  \begin{center}
    \includegraphics[width=0.7\textwidth]{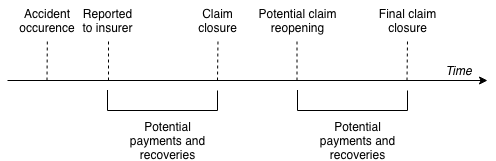}
  \end{center}
  \caption{Development of a claim}
  \label{fig:claimrunoff}
\end{figure}

Figure \ref{fig:claimrunoff} shows a diagram of the development of a typical
claim. We first point out that there can be a time difference, known as the 
reporting lag, between an accident's occurrence and its reporting. The accidents
which have been reported to the insurer, but not yet settled, are known as 
reported but not settled (RBNS) or, equivalently, incurred but not enough 
reported (IBNER) claims, while the accidents which have occurred but are yet
unknown to the insurer are known as incurred but not reported (IBNR) claims. The
reserving actuary is interested in estimating the ultimate loss amounts 
associated with accidents that have already occurred. As demonstrated in Figure
\ref{fig:claimrunoff}, it is possible for a closed claim to re-open, and it is
also possible for a claim to be closed without any cash flows.

In this paper, we are concerned with RBNS/IBNER, but not IBNR, claims, due to 
limitations of available data, as for IBNR claims we do not have individual 
claim feature information available. Also, the claims we study encompass closed 
claims to allow for the possibility of claim re-opening.

\subsection{Bayesian Inference on Neural Networks} \label{section:bnn}

In this section, we briefly discuss Bayesian inference on neural networks, 
following the discourse in \cite{gravesPracticalVariational2011} and 
\cite{blundellWeightUncertainty2015}.

Consider a neural network parameterized by weights $w$ that takes input $x$ and
returns output $y$. We can view this general formulation as a probabilistic 
model $P(y|x, w)$; as an example, in the case of linear regression where $y \in
\mathbb{R}$, mean squared loss is specified, and constant variance is assumed, 
$P(y|x, w)$ corresponds to a Gaussian distribution.

Rather than treating $w$ as a fixed unknown parameter, we adapt a Bayesian
perspective and treat $w$ as a random variable with some prior distribution 
$P(w)$. The task is then to compute the posterior distribution $P(w|x, y)$ given
the training data. We can then calculate the posterior predictive distribution
$P(y^*|x^*) = \mathbb{E}_{P(w|x, y)}[P(y^*|x^*, w)]$, where $x^*$ is a new data
point to be scored and $Y^*$ the corresponding unknown response. However, 
determining $P(w|x, y)$ analytically is intractable, and convergence of Markov
chain Monte Carlo (MCMC) to the actual posterior for nontrivial neural networks 
is too slow to be feasible.
Variational inference provides a workaround to this problem by approximating the
posterior with a more tractable distribution $q(w|\theta)$, which is often 
chosen to come from the mean-field family, i.e., $q(z) = \prod_i q(z_i)$, but can
be more general \cite{blundellWeightUncertainty2015}. We then formulate an 
optimization problem to find the parameters $\theta$. Specifically, we minimize
the Kullback-Leibler (KL) divergence from the true posterior distribution $P$ to
the approximate distribution $q$. KL divergence is defined in general for 
probability distributions $P$ to $Q$ as

\begin{equation}
    D_{KL}(Q \Vert P) = \int q(x) \log \frac{q(x)}{p(x)} dx.
\end{equation}

Our optimization problem can then be stated as

\begin{align}
    \theta^* &= \arg \min_\theta D_{KL}(q(w|\theta) \Vert P(w|\mathcal{D}))\\
             &= \arg \min_\theta \int q(w|\theta) \log \frac{q(w|\theta)}{P(w|\mathcal{D})} dw\\
             &= \arg \min_\theta \int q(w|\theta) \log \frac{q(w|\theta)}{P(w)P(\mathcal{D}|w)} dw \label{eq:2}\\
             &= \arg \min_\theta D_{KL}(q(w|\theta) \Vert P(w)) - \mathbb{E}_{q(w|\theta)}[\log P(\mathcal{D}|w)] \label{eq:1}
\end{align}

where $\mathcal{D} = (x_i, y_i)_i$ is the training dataset. We remark that, in Equation \ref{eq:2}, the $P(\mathcal{D})$ term resulting from applying Bayes' theorem to $P(w|\mathcal{D})$ disappears because it is irrelevant to the optimization.

Equation \ref{eq:1} then gives us the optimization objective\footnote{We note that $-\mathcal{F}(\mathcal{D}, \theta)$ is often referred to as the evidence lower bound (ELBO) in the machine learning literature.}

\begin{equation}
    \mathcal{F}(\mathcal{D}, \theta) = D_{KL}(q(w|\theta) \Vert P(w)) - \mathbb{E}_{q(w|\theta)}[\log P(\mathcal{D}|w)].
\end{equation}

In practice, during
training, where $\mathcal{D}$ is randomly split into $M$ mini-batches
$\mathcal{D}_1,\dots,\mathcal{D}_M$, we compute 
$\mathcal{F}(\mathcal{D}, \theta) = \sum_1^M \mathcal{F}_i(\mathcal{D}_i, \theta)$, 
where

\begin{equation}
    \mathcal{F}_i(\mathcal{D}_i, \theta) = \frac{|\mathcal{D}_i|}{|\mathcal{D}|}D_{KL}(q(w|\theta) \Vert P(w)) - \mathbb{E}_{q(w|\theta)}[\log P(\mathcal{D}_i|w)],
\end{equation}

which we approximate with

\begin{equation}
    \mathcal{F}_i(\mathcal{D}_i, \theta) \approx \sum_{j=1}^{|\mathcal{D}_i|} \frac{1}{|\mathcal{D}|} ( \log q(w^{(j)}|\theta) - \log P(w^{(j)}) ) - \log P(\mathcal{D}_i|w^{(j)}),
\end{equation}

where the $w^{(j)}$ are sampled independently from $q(w|\theta)$. In other
words, we sample from the weights distribution just once for each training 
sample.

\subsection{Mixture Density Networks}\label{section:mdn}

In practical applications, the response variables we try to predict often have
multimodal distributions and exhibit heteroscedastic errors. This is particular 
relevant in forecasting claims cash flows since, in a given time period, there 
could be a large payment or little or no payment. An MDN allows the output to
follow a mixture of arbitrary distributions and estimate each of its parameters 
with the neural network. Recall that a mixture distribution has a distribution 
function of the form

\begin{equation}
    F(z) = \sum_{i = 1}^n w_i P_i(z),
\end{equation}

where each $w_i \geq 0$, $\sum w_i = 1$, and $n$ is the number of component 
distributions $P_i$. Letting $\mathcal{P}$ denote the union of the sets of 
parameters for the distributions $P_1, \dots, P_n$, the neural network must then
output $n + |\mathcal{P}|$ values. The first $n$ values determine the 
categorical distribution of the mixing weights, and the $|\mathcal{P}|$ outputs
parameterize the component distributions.

By obtaining distributions rather than single points as prediction outputs, we
also gain a straightforward mechanism to quantify the uncertainty of individual
cash flow forecasts.

We emphasize now the difference between the uncertainty captured by specifying a
distribution as the neural network's output, as discussed here, and the 
uncertainty in the weight distribution, as discussed in Section 
\ref{section:bnn}. The former corresponds to the irreducible pure randomness,
while the latter reflects uncertainty in parameter estimation.

\section{Data and Model}

In this section, we describe the data used for our experiments and the proposed
model. The dataset used and relevant code are available on 
GitHub.\footnote{\url{https://github.com/kasaai/bnn-claims}} The experiments are
implemented using the R programming language 
\cite{rdevelopmentcoreteamLanguageEnvironment2011} and TensorFlow 
\cite{tensorflow2015-whitepaper}.

\subsection{Data}

We utilize the individual claims history simulator of 
\cite{gabrielliIndividualClaims2018} to generate data for our experiments. For 
each claim, we have the following static information: line of business, labor 
sector of the injured, accident year, accident quarter, age of the injured, part
of body injured, and the reporting year. In addition, we also have 12 years of 
claim development information, in the form of cash flows and claims statuses 
(whether the claim is open or not). The generated claims history exhibit 
behaviors such as negative cash flows for recoveries and late reporting, which 
mimic realistic scenarios.

Since we only have claims data and not policy level and exposure data, we study 
only reported claims.

\subsection{Experiment Setup}

We first simulate approximately 500,000\footnote{The number of claims generated
is stochastic; in our case, we draw 500,904 claims.} claims using the simulator,
which provides us with the full development history of these claims. The claims 
cover accident years 1994 to 2005. Since we are concerned with reported claims,
we remove claims with report date after 2005, which leaves us with N = 497,516 
claims. In this paper, we assume that each claim is fully developed at 
development year 11 (note that in the dataset the first development year is 
denoted year 0). More formally, we can represent the dataset as the collection

\begin{equation}
    \mathcal{D} = \{ (X^{(j)}, (C_i^{(j)})_{0 \leq i \leq 11}, (S_i^{(j)})_{0 \leq i \leq 11}): j\in {1,\dots, N}\},
\end{equation}

where $X$, $(C_i)$, and $(S_i)$ denote the static claim features, incremental 
cash flow sequences, and claim status sequences, respectively, and $j$ indexes 
the claims.

To create the training and testing sets, we select year 2005 as the evaluation 
year cutoff. For the training set, any cash flow information available after
2005 is removed. In symbols, we have

\begin{equation}
    \mathcal{D}_{\text{train}} = \{ (X^{(j)}, (C_i^{(j)}), (S_i^{(j)})): i + \text{AY}^{(j)} \leq 2005, j \in {1,\dots, N}\},
\end{equation}
where $\text{AY}^{(j)}$ denotes the accident year associated with claim $j$.

For each claim in the training set, we create a training sample for each time 
period after development year 0. The response variable consists of cash flow 
information available as of the end of each time period until the evaluation 
year cutoff, and predictors are derived from information available before the
time period. For a claim $j$, we have the following input-output pairs:

\begin{equation}\label{eq:training-samples}
    \{((X^{(j)}, (C_0^{(j)}, \dots, C_i^{(j)}), (S_0^{(j)}, \dots, S_i^{(j)})), (C_{i+1}^{(j)}, \dots, C_{k^{(j)}}^{(j)})): i = 0, \dots, k^{(j)} - 1\},
\end{equation}

where $k^{(j)}$ denotes the latest development year for which data is available
for claim $j$ in the training set. As an example, if a claim has an accident 
year of 2000, five training samples are created. The first training sample has 
cash flows from 2001 to 2005 for the response and one cash flow value from 2000
for the predictor, while the last training sample has only the cash flow in year
2005 for the response and cash flows from 2000 to 2004 for the predictor.

We note here that we do not predict future claim statuses. As we will discuss in
Section \ref{section:outputdist}, our output distributions, which contain point 
masses at zero, can accommodate behaviors of both open and closed claims.

The training samples in Equation \ref{eq:training-samples} undergo additional 
transformations before they are used for training our model. We discuss these 
transformations in detail in the next section.

\subsection{Feature Engineering} \label{sec:fe}

We exhibit the predictors used and associated transformations in Table 
\ref{table:vars}. In the raw simulated data, each time period has a cash flow
amount along with a claim open status indicator associated with it. We take the
cash flow value and derive two variables from it: the paid loss and the 
recovery, corresponding to the positive part and negative part of the cash flow,
respectively. In other words, for each claim and for each time step, at most one
of paid loss and recovery can be nonzero. For predictors, both the paid losses 
and recoveries are centered and scaled with respect to all instances of their 
values in the training set. The response cash flow values are not normalized.

Claim status indicator, which is a scalar value of 0 or 1, is one-hot encoded
(i.e., represented using dummy variables);
for each training sample, a claim status value is available for each time step.

Development year is defined as the number of years since the accident occurred. 
It is then scaled to $[0, 1]$.

The static claim characteristic variables include age, line of business, 
occupation of the claimant, and the injured body part. Of these, age is numeric
while the others are categorical. We center and scale the age variable and 
integer-index the others, which are fed into embedding layers
\cite{guoEntityEmbeddings2016}, discussed in the next section.

As noted in the previous section, the response variable and the sequence 
predictors can have different lengths (in terms of time steps) from one sample
to the next. To facilitate computation, we pre-pad and post-pad the sequences 
with a predetermined masking value (we use 99999) so that all sequences have a 
fixed length of 11.

\begin{table}[ht]
\centering
\begin{tabular}[t]{lll}
\toprule
Variable & Type & Preprocessing\\
\midrule
Paid loss history      & Numeric sequence         & Centered and scaled\\
Recovery history       & Numeric sequence         & Centered and scaled\\
Claim status history   & Categorical sequence     & One-hot encoded\\
Development year       & Integer                  & Scaled to [0, 1]\\
Age                    & Numeric                  & Centered and scaled\\
Line of business       & Categorical              & Indexed\\
Claim code (occupation)& Categorical              & Indexed\\
Injured part           & Categorical              & Indexed\\
\bottomrule\\[1ex]
\end{tabular}
\caption{Predictor variable and transformations.}
\label{table:vars}
\end{table}%

\subsection{Claims Forecasting Model}\label{sec:claims-forecasting-model}

\begin{figure}
  \begin{center}
    \includegraphics[width=0.5\textwidth]{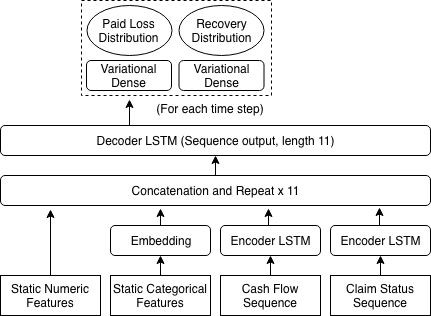}
  \end{center}
  \caption{Claims forecasting model architecture. Note that the weights for the 
  variational dense layers are shared across time steps.}
  \label{fig:architecture}
\end{figure}

We utilize an encoder-decoder architecture with sequence output similar to the 
model proposed by \cite{kuo2019deeptriangle}. The architecture is 
illustrated in Figure \ref{fig:architecture}. We first provide a brief overview
of the architecture, then provide details on specific components later in the 
section.

The output in our case is the future sequence of distributions of loss payments,
and the input is the cash flow and claim status history along with static claim 
characteristics.

Static categorical variable inputs, such as labor sector of the injured, are 
indexed, then connected to embedding layers \cite{guoEntityEmbeddings2016} and 
sequential data, including payments and claim statuses, are connected to long 
short-term memory (LSTM) layers \cite{hochreiterLongShortterm1997}.

The encoded inputs are concatenated together, then repeated 11 times before being
passed to a decoder LSTM layer which returns a sequence. The length of this 
output sequence is so chosen to match our requirement to forecast a maximum of 
11 steps into the future. Each time step of this sequence is connected to two 
dense variational layers, each of which parameterizes an output distribution, 
corresponding to paid losses and recoveries, respectively. The weights of the 
dense variational layer (for paid loss and recovery) are each shared across the
time steps. In other words, for each training sample, we output two 
11-dimensional random variables, each of which can be considered as a collection
of 11 independent but non-identically distributed random variables.

We use the same loss function for each output and weight them equally for model
optimization. The loss function is the negative log-likelihood given the target 
data with an adjustment for variable output lengths which we discuss in detail 
in Section \ref{sec:loss-function}.

In the remainder of this section, we discuss in more detail embedding layers, 
LSTM, our choices of the variational and distribution layers, the loss function,
and model training.

\subsubsection{Embedding Layer}

An embedding layer maps each level of a categorical variable to a fixed-length 
vector in $n$-dimensional Euclidean space. The value of $n$ is a hyperparameter 
chosen by the modeler; in our case we select $n = 2$ for all embedding layers, 
which means we map each factor level to a point in $\mathbb{R}^2$. In contrast 
to data-preprocessing dimensionality techniques such as principal component 
analysis (PCA) or t-distributed stochastic neighbor embedding (t-SNE), the 
values of the embeddings are learned during training of the neural network.

\subsubsection{Long Short-Term Memory}

LSTM is a type of recurrent neural network (RNN) architecture that is 
appropriate for sequential data, such as natural language or time series (our 
use case). Empirically, LSTM is more apt at handling data with longer term 
dependencies than simple RNN \cite{lecunDeepLearning2015}. In our model, for 
both of the sequence input encoders and the decoder, we utilize a single layer 
of LSTM with 3 hidden units. We found that these relatively small modules 
provided reasonable results, but did not perform comprehensive tuning to test
deeper and larger architectures. 

\subsubsection{Dense Variational Layer}

We choose Gaussians for both the prior and surrogate posterior distributions 
over the weights of the dense layers. The prior distribution is constrained to 
have zero mean and unit variance while for the surrogate posterior both the mean and
variance are trainable. Symbolically, if we let $w$ denote the weights 
associated with each dense layer, we have

\begin{equation}
    P(w) = \mathcal{N}(0, I) \text{ and}
\end{equation}
\begin{equation}
    q(w|\theta) = \mathcal{N}(\mu(\theta), \sigma(\theta)),
\end{equation}

where $\theta$ represents trainable parameters. The KL term associated with
estimation, as described in Section \ref{section:bnn}, is added to the neural 
network loss during optimization. The dense layers each output four units to 
parameterize the output distributions which we discuss next.

\subsubsection{Output Distribution}\label{section:outputdist}

For each of the 11 time steps we forecast, we parameterize two distributions, 
one for paid losses and one for recoveries. Each of the distributions is chosen 
to be a mixture of a shifted log-normal distribution and a deterministic 
degenerate distribution localized at zero, representing cash flow and no cash 
flow, respectively. Denoting $\{v_1,\dots,v_4\}$ to be the output of the 
preceding dense variational layer, we have the following as the distribution 
function:

\begin{equation} \label{eq:fx}
    F(y) = w_1(v_1, v_2)P(y; v_3, (\alpha + \lambda\sigma(\beta v_4))^2) + w_2(v_1, v_2)F_0(y).
\end{equation}

We now describe the components of Equation \ref{eq:fx}. First, 

\begin{equation}
    w_i = \frac{e^{v_i}}{e^{v_1} + e^{v_2}}\quad (i = 1, 2)
\end{equation}

are the normalized mixing weights. $P(y; \mu, \sigma^2)$ denotes the 
distribution function for a log-normal distribution with location and scale 
parameters $\mu$ and $\sigma$, respectively, shifted to the left by 0.001 to 
accommodate zeros in the data. In other words, if $Y \sim LN(\mu, \sigma^2)$, 
then $Y - 0.001 \sim P$. This modification is necessary because we have zero 
cash flows in the response sequences, and computing the likelihood becomes
problematic since the support of the log-normal distribution does not contain
zero.\footnote{As of the writing of this paper, TensorFlow Probability attempts
to evaluate all terms in the likelihood of the mixture.}

\begin{equation}
    F_{0}(y)=\left\{\begin{matrix} 1, & \mbox{if }y\ge 0 \\ 0, & \mbox{if }y<0 \end{matrix}\right.
\end{equation}

is the distribution function for the deterministic distribution, and $\sigma$ 
is the sigmoid function defined as

\begin{equation}
    \sigma(t) = \frac{1}{1 + e^{-t}}.
\end{equation}

In Equation \ref{eq:fx}, the constants $\alpha, \beta > 0$ are included for 
numerical stability and can be tuned as hyperparameters, although we set them 
to $0.001$ and $0.01$, respectively, for our experiments. The purpose of 
$\lambda$ is also numerical stability, as the scale parameter of the log-normal
distribution is hard to learn in practice, so we bound it above by a constant. 
In our experiments we fix it to be $0.7$.

The net loss prediction for each development year is then the difference of the 
paid loss and recovery amounts. We assume that the output distributions, 
conditional on their parameters, are independent across the time steps, which 
facilitates the derivation of the loss function in the next section. We remark 
that this is a weakness of our approach, since the independence assumption is
not fulfilled in practice.

\subsubsection{Loss Function}\label{sec:loss-function}

We calculate the log-likelihood loss for each output by computing the log 
probability of the true labels with respect to the output distributions. As
mentioned in Section \ref{sec:fe}, the output sequences may contain masking 
values that need to be adjusted. Specifically, we marginalize out the components
with the masking values. Formally, for a given training sample, let $M$ denote 
the masking constant, $Y=(Y_1,\dots,Y_{11})$ denote a single output sequence,
$Y_{\mathcal{T}_M} = (Y_i: Y_i = M)$, and, abusing notation, also let 
$\mathcal{T}_M = \{i: Y_i = M\}$. Then,

\begin{align}
    f_{Y\setminus Y_{\mathcal{T}_M}}(Y) &= \int_{Y_{\mathcal{T}_M}} f(Y) dY_{\mathcal{T}_M}\\
         &= \int_{Y_{\mathcal{T}_M}} f(Y\setminus Y_{\mathcal{T}_M}, Y_{\mathcal{T}_M}) dY_{\mathcal{T}_M}\\
         &= f_{Y\setminus Y_{\mathcal{T}_M}}(Y\setminus Y_{\mathcal{T}_M}) \int_{Y_{\mathcal{T}_M}} f_{Y_{\mathcal{T}_M}}(Y_{\mathcal{T}_M}) dY_{\mathcal{T}_M} \quad \text{(by independence)}\\
         &= f_{Y\setminus Y_{\mathcal{T}_M}}(Y\setminus Y_{\mathcal{T}_M})\\
         &= \prod_{i \notin \mathcal{T}_M} f_{Y_i}(Y_i) \quad \text{(by independence)}.
\end{align}

The adjusted log-likelihood for a single training sample then becomes

\begin{align}
    \log\mathcal{L}(Y) &= \log  f_{Y\setminus Y_{\mathcal{T}_M}}(Y)\\
                       &= \log \prod_{i \notin \mathcal{T}_M} f_{Y_i}(Y_i)\\
                       &= \sum_{i = 1}^{T}\log f_{Y_i}(Y_i),
\end{align}

where $T$ is $\min \mathcal{T}_M - 1$ if $\mathcal{T}_M$ is nonempty and 11 
otherwise.

Hence, the log-likelihood associated with a training sample is the sum of the 
log probabilities of the non-masked elements. The negative log-likelihood, 
summed with the KL term associated with the variational layers, discussed in 
Section  \ref{section:bnn}, comprise the total network loss for optimization. 
The contribution of a single training sample to the loss is then

\begin{equation}
    \sum_{i = 1}^{T}\log f_{Y_i}(Y_i) + \frac{1}{|\mathcal{D}|}\sum_{k=1}^2D_{KL}(q_k(w_k|\theta_k) \Vert P_k(w_k)),
\end{equation}

where $k=1, 2$ correspond to variational layers for parameterizing the paid loss
and recovery distributions, respectively.

\subsubsection{Training and Scoring}

We use a random subset of the training set consisting of 5\% of the records as 
the validation set for determining early stopping and scheduling the learning 
rate. For optimizing the neural network, we use stochastic gradient descent with
an initial learning rate of 0.01 and a minibatch size of 100,000, and halve the 
learning rate when there is no improvement in the validation loss for five
epochs. Training is stopped early when the validation loss does not improve for 
ten epochs; we cap the maximum number of epochs at 100. Here, an \textit{epoch} 
refers to a complete iteration through our training dataset, and the 
\textit{minibatch} size refers to how many training samples we randomly sample 
for each gradient descent step.

To forecast an individual claim, we construct a scoring data point by using all
data available to us as of the evaluation date cutoff. In other words, the last
elements of the predictor cash flow sequences correspond to the actual value
from the cutoff year. Hence, in the output sequence, the first element 
corresponds to the prediction of the cash flow distribution of the year after
the cutoff.

Recall that our model weights are a random variable, so each time we score a new
data point, we are sampling from the distribution of model weights and expect to
obtain different parameters for the distributions of cash flows. The output 
distributions themselves are also stochastic and can be sampled from. Following
\cite{kendallWhatUncertainties2017}, we refer to the former variability as 
\textit{epistemic} uncertainty and the latter as \textit{aleatoric} uncertainty.
Epistemic uncertainty reflects uncertainty about our model which can be reduced
given more training data, while aleatoric uncertainty reflects the irreducible 
noise in the observations. In actuarial science literature, these concepts are 
also known as parameter estimation uncertainty and process uncertainty, 
respectively \cite{wuthrichNonlifeInsurance2017}.

Generating a distribution of future paths of cash flows amounts to repeating the
procedure of drawing a model from its distribution, calculating the output 
distribution parameters using the drawn model, then drawing a sample of cash 
flows from the distribution. Since we decompose the cash flows into paid losses 
and recoveries, we can obtain net amounts by subtracting the generated 
recoveries from the generated paid losses.

If we were only interested in point estimates, we can avoid sampling the output
distributions and simply compute their means, since given the model weights we
have closed form expressions for the distributions given by Equation 
\ref{eq:fx}. Note that we would still have to sample the model weights.

\section {Results and Discussion}

We evaluate our modeling framework by qualitatively inspecting samples paths 
generated for an individual claim and also comparing the aggregate estimate of 
unpaid claims with a chain ladder estimate. We note that, to the best of our 
knowledge, prior to the current paper, there is not a benchmark for individual
claims forecasts. We also discuss the extensibility of our approach to 
accommodate company specific data elements and expert knowledge.

\subsection{Individual Claim Forecasts}

\begin{figure}
  \begin{center}
    \includegraphics[width=0.8\textwidth]{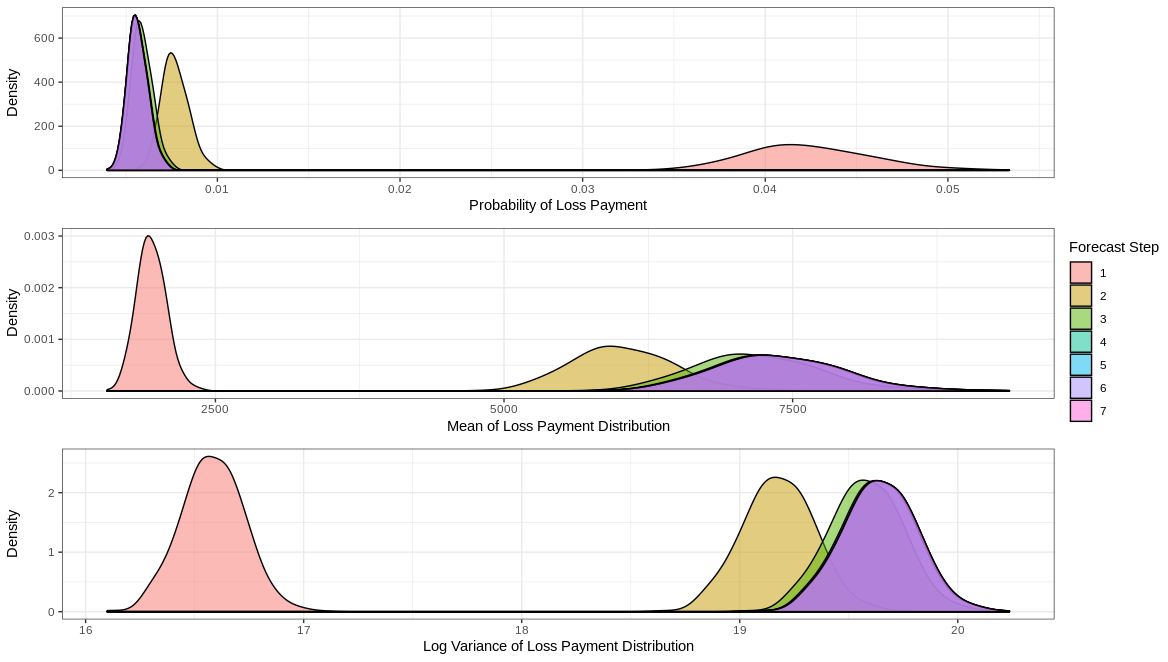}
  \end{center}
  \caption{Posterior distributions for a single claim at development year 4. 
  (Top) Payment probability. (Middle) Mean of loss payment distribution. 
  (Bottom) Log variance of payment distribution.}
  \label{fig:prob_cash_flows}
\end{figure}

Figure \ref{fig:prob_cash_flows} shows various posterior densities, obtained by
sampling the model weights 1,000 times, of parameters for the output 
distributions of a single claim. Our model assigns a higher probability of 
payment in the next year along with more variability around that probability. 
It can be seen that the expected probability of a payment decreases as we 
forecast further into the future. Given that a loss payment occurs, we see from
the middle and bottom plots that both the expected value and the variability for
both the mean and variance of the log-normal distributions increase with time. 
In other words, for this particular claim, loss payments become less likely as 
time goes on, but if they occur, they tend to be more severe and the model is 
less certain about the severity distribution.

\begin{figure}
  \begin{center}
    \includegraphics[width=0.8\textwidth]{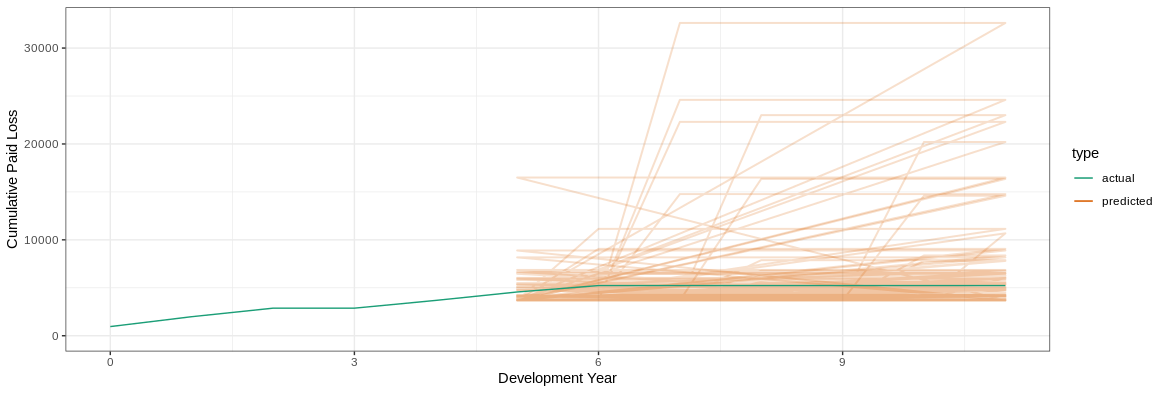}
  \end{center}
  \caption{1,000 samples of cumulative cash flow paths for a single claim. The 
  training data includes development year 4 and the predictions begin in development year 5.}
  \label{fig:claim_paths}
\end{figure}

In Figure \ref{fig:claim_paths} we show plausible future developments of the
claim. Note that one thousand samples are drawn, so the handful of scenarios that
present large payments represents a small portion of the paths, which is 
consistent with the distributions of parameters we see above.

\subsection{Aggregate estimates}

To compute a point estimate for the total unpaid losses, we score each claim 
once (i.e., sample from the weights distribution once) to obtain a distribution 
for each time step for paid losses and recoveries. We then compute the means of
the paid loss and recovery distributions and take the differences to obtain the 
net cash flow amount. The final unpaid loss estimate is then the sum of all the 
net amounts up to and including development year 11. Since there is randomness 
in the neural network weight initialization, we instantiate and train the model
10 times, and take the average of the predicted future paid losses across the 
10-model ensemble. In practice, one could distribute the prediction procedure 
over a computing cluster and draw more samples for an even more stable estimate.

For the chain ladder development method benchmark, we aggregate the data into a 
report year triangle (to exclude IBNR), calculate ultimate losses using all-year
weighted age-to-age factors, then subtract the already paid amounts to arrive at
the unpaid estimate. The unpaid amounts from the two approaches are then 
compared to the actual unpaid amounts. The results are shown in Table
\ref{table:results}.

\begin{table}[ht]
\centering
\begin{tabular}[t]{lll}
\toprule
Model & Forecast & Error\\
\midrule
Actual                                & 109,216,388  & --\\
Chain Ladder                          & 108,040,572  & $-1.08\%$\\
Ours (10-Model Ensemble)              & 102,502,710  & $-6.15\%$\\
\bottomrule\\[1ex]
\end{tabular}
\caption{Aggregate forecast results.}
\label{table:results}
\end{table}%

While we are not improving on chain ladder estimates at the aggregate level for 
this particular simulated dataset, our approach is able to provide individual 
claims forecasts, which are interesting in their own right.

\subsection{Extensibility}

One of the primary advantages of neural networks is their flexibility. 
Considering the architecture described in Section 
\ref{sec:claims-forecasting-model}, we can include additional predictors by 
appending additional input layers. These inputs can be unstructured and we can 
leverage appropriate techniques to process them before combining with the 
existing features. For example, we may utilize convolutional layers to transform
image inputs and recurrent layers to transform audio or text inputs.

The forms of the output distributions can also be customized. We choose a 
log-normal mixture for our dataset, but depending on the specific book of 
business, one may want to specify a different distribution, such as a Gamma or a
Gaussian. As we have done in constraining the scale parameter of the log-normal 
distribution, the modeler also has the ability to bound or fix specific 
parameters as situations call for them.

\section{Conclusion}

We have introduced a framework for individual claims forecasting that can be 
utilized in loss reserving. By leveraging Bayesian neural networks and 
stochastic output layers, our approach provides ways to learn uncertainty from 
the data. It is also able to produce cash flow estimates for multiple future
time periods. Experiments confirm that the approach gives reasonable results and
provides a viable candidate for future work on individual claims forecasting to 
benchmark against.

There are a few potential avenues for enhancements and extensions, including 
larger scale simulations to evaluate the quality of uncertainty estimates, 
evaluating against new datasets, and relaxing simplifying assumptions, such as 
the independence of time steps to obtain more internally consistent forecasts.
One main limitation of the proposed approach and experiments is that we focus on
reported claims only. With access to policy level data, a particularly interesting 
direction of research would be to extend the approach to also estimate IBNR claims.

\subsubsection*{Acknowledgments}

We thank Sigrid Keydana, Daniel Falbel, Mario Wüthrich, and volunteers from the
Casualty Actuarial Society (CAS) for helpful discussions. This work is supported
by the CAS.

\bibliographystyle{abbrv}
\bibliography{bnn-claims}  






%
%
%

\end{document}